\documentclass[twocolumn,showpacs]{revtex4}
\usepackage{dcolumn}
\usepackage{amsmath}
\usepackage[xdvi]{graphicx}

\newcommand{\bra}{\left\langle}
\newcommand{\ket}{\right\rangle}

\begin{document}

\title{Temperature of a Hamiltonian system given as the 
effective temperature \\ 
of a non-equilibrium steady state  Langevin 
thermostat
}
\author{Kumiko Hayashi and Mitsunori Takano} 
\affiliation
{Department of Physics, Waseda University, Tokyo 169-8555, Japan}

\date{\today}

\begin{abstract}
In non-equilibrium steady states (NESS) far from equilibrium, it is 
known that the Einstein relation is violated. Then, the ratio of the 
diffusion coefficient to the mobility is called an effective temperature,  
and the physical relevance of this effective temperature has been studied 
in several works. Although the physical relevance is not yet completely 
clear, it has been found that the role of an effective temperature in 
NESS is indeed analogous to that of the temperature in equilibrium 
systems in a number of respects. In this paper, we find further evidence 
establishing this analogy. We employ a non-equilibrium Langevin system 
as a thermostat for a Hamiltonian system and find that the kinetic 
temperature of this Hamiltonian system is equal to the effective 
temperature of the thermostat. 
\end{abstract}

\maketitle


Fluctuation-dissipation relations (FDRs) relate dynamical properties of 
fluctuations in systems under equilibrium conditions to linear transport 
properties of non-equilibrium systems through the detailed-balance 
condition  \cite{lrt}.  Representative examples of FDRs are the Einstein 
relation, which relates a diffusion coefficient and a mobility, and 
the Green-Kubo relation, which relates current fluctuations and the  
corresponding conductivities. 

In recent years, the properties of fluctuations and linear responses 
to perturbations have been investigated for non-equilibrium states even 
outside the linear response regime, specifically in steady state systems 
\cite{arr,sho,bl} and in aging systems \cite{ckp,crisan,barrat,takano}. 
Although, we cannot expect FDRs to be generally valid outside the 
linear response regime, there have been several relations proposed and 
investigated recently that represent extensions of FDRs to systems far 
from equilibrium \cite{hs3a,hs5,harada,toyabe}.

In Refs. \cite{hs3a,hs5}, the violation of FDRs is studied in the case of 
a non-equilibrium one-dimensional Langevin system in which  a Brownian 
particle is subject to a spatially constant driving force $f$ and a 
periodic potential $U(x)$. Explicitly, the system studied there is 
\begin{eqnarray}
&& \gamma \dot{x} = -\frac{\partial U(x)}{\partial x} + f 
+\xi(t), \nonumber \\
&& \bra \xi(t)\xi(t')\ket = 2\gamma T\delta (t-t'),  
\label{model0}  
\end{eqnarray}
where $x(t)$ is the position of the Brownian particle, $\xi(t)$ is Gaussian 
noise, $\gamma$ is the friction coefficient, and $T$ is the temperature of 
the environment. (The Boltzmann constant is set to unity.)  In this model, 
in the linear response regime, the Einstein relation  
\begin{figure}
\begin{center}
\includegraphics[width=6.0cm]{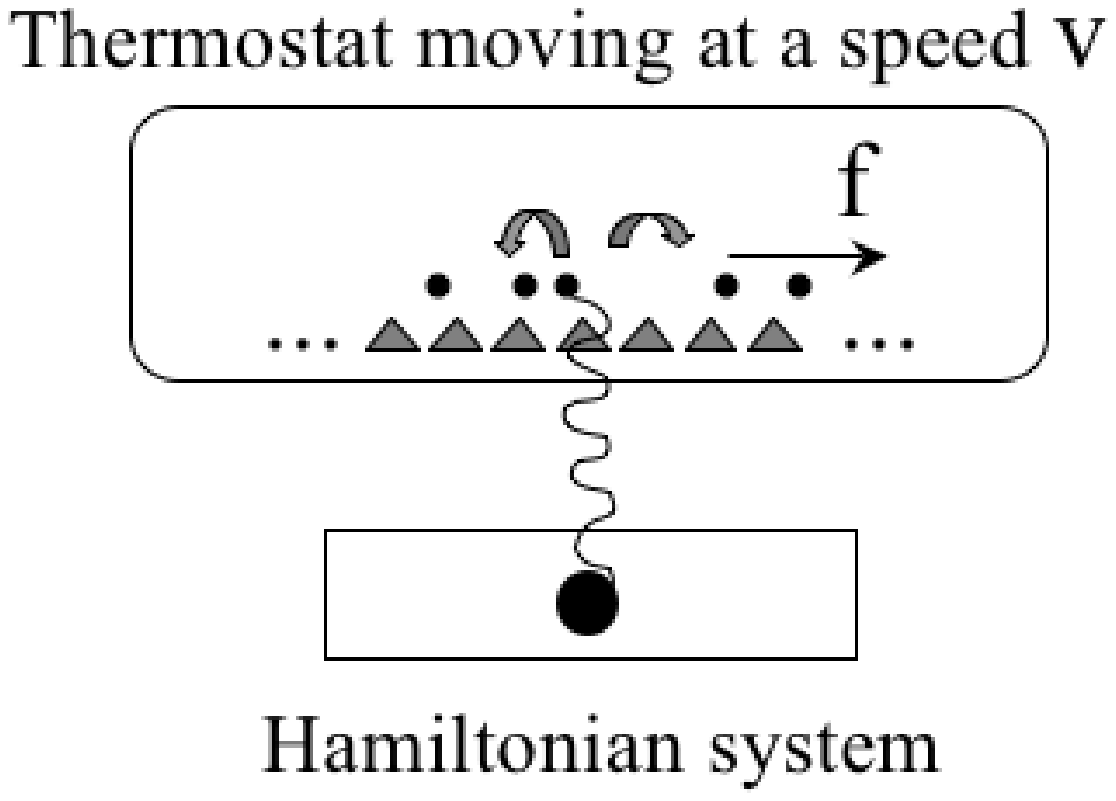}
\caption{Schematic depiction of the ``moving thermostat.''  The 
thermostat moves at a speed $v$ relative to the Hamiltonian system. 
}
\label{fig:ondo1}
\end{center}
\end{figure}
\begin{equation}
D=\mu_{\rm d}T
\label{ein}
\end{equation}
holds, where $D$ is the diffusion coefficient and $\mu_{\rm d}$ is the 
differential mobility defined as   
\begin{eqnarray}
&&D(f)\equiv
\lim_{t\to\infty} \frac{\bra (x(t)-x(0)-v_{\rm s}(f)t)^2\ket}{2t}, \\ 
&&\mu_{\rm d}(f)\equiv\frac{{\rm d}v_{\rm s}(f)}{{\rm d}f}.  
\end{eqnarray}
Here $v_{\rm s}(f)$ is the steady state velocity of the Brownian particle 
and is known as the Stratonovich formula \cite{hs3a,risken}.  However,  
outside the linear response regime, i.e. for 
large $f$,  the above Einstein relation does not hold.  In such 
situations, as an extension of the concept of temperature, it is natural 
to define the following quantity:
\begin{equation}
\Theta(f)\equiv \frac{D(f)}{\mu_{\rm d}(f)}.  
\label{the}
\end{equation}
Then, outside the linear response regime, we have $\Theta\ne T$.  
Thus, the introduction of $\Theta$ allows us to define an extended 
Einstein relation that applies to non-equilibrium steady states (NESS) 
far from equilibrium, although this leads to the question of the physical 
significance of $\Theta$. In Refs. \cite{hs3a,hs5},  in order to elucidate 
the physical significance of $\Theta$, a large-scale  description of the 
system was derived by applying a perturbation method to the Fokker-Planck 
equation \cite{hs3a} and by considering a finite time average of the 
Langevin equation \cite{hs5}.  With these treatments, it was found that 
$\Theta$ plays the role of a temperature in the large-scale description of 
the non-equilibrium Langevin system (\ref{model0}), and for this reason, 
it is referred to as an effective temperature.

In this paper, we present a study that further establishes the role of 
$\Theta$ as an effective temperature for NESS. Here, we employ a Langevin 
system in a NESS as a thermostat for a Hamiltonian system, and we 
investigate the temperature of the Hamiltonian system established by 
this thermostat.  More precisely, we set out to determine whether the 
kinetic temperature  of the Hamiltonian system is equal to $\Theta$ in 
this situation.  We find that in fact the kinetic temperature is equal to  
$\Theta$ in the case that the thermostat moves at the speed 
$v=-v_{\rm s}(f)$, so that the average velocity of the Brownian particle 
relative to the Hamiltonian system is zero. (See the schematic in Fig. 
\ref{fig:ondo1} and {\it Moving thermostat} for details.)


%
\begin{figure}
\begin{center}
\includegraphics[width=7cm]{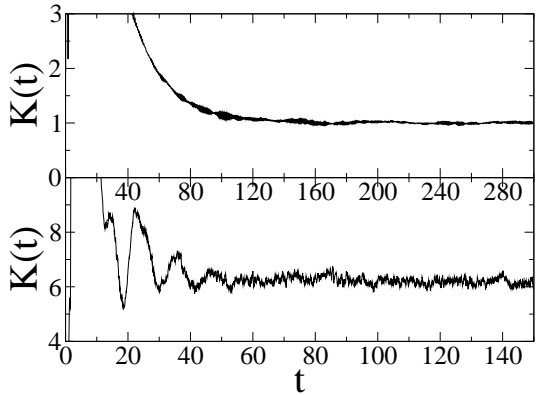}
\caption{$K(t)$ obtained using a stationary thermostat  as 
a function  of time in the cases $f=0$ (upper) and  $f=10$ (lower).  
These results were obtained from $5000$ samples. 
}
\label{fig:ondo2}
\end{center}
\end{figure}

{\it Stationary thermostat}.  
In Fig. \ref{fig:ondo1}, we present a schematic depiction of the 
model we study. In this section, we consider a combined system 
consisting of a Hamiltonian system in contact with a Langevin thermostat 
in the case that the two are relatively at rest, i.e. $v=0$. 

The Langevin thermostat consists of $N$ Brownian particles ($N=20$) 
which are confined to move along a single direction, say the $x$ direction.  
Because there is no interaction between the Brownian particles, the 
statistical properties of each are the same as those in the 
model (\ref{model0}), studied in Refs. \cite{hs3a,hs5}.  Each Brownian  
particle is subject to a constant driving force $f$ and a periodic 
potential $U(x_i)=(U_0/T)\sin(2\pi x_i/\ell)$, where $x_i$ represents the 
position of the $i$-th particle. The size of the thermostat is chosen as  
$20\ell$ ($-10\ell\le x_i \le 10\ell$), and periodic boundary conditions 
are  imposed on the Brownian particles. 

The Hamiltonian system we consider is a one-dimensional system consisting 
of a single particle. Each Brownian particle in the Langevin system 
interacts with this Hamiltonian particle through the potential 
$U_{\rm int}=\varepsilon (x_i-x^{\rm H})^2/2$ for $|x_i-x^{\rm H}| < r_{\rm c}$ 
and $U_{\rm int}=0$ otherwise, where $x^{\rm H}$ is the position of the 
Hamiltonian particle, and $r_{\rm c}$ is the cut-off length of the 
interaction. This particle is confined to  the region 
$-5\ell \le x^{\rm H}\le 5\ell$ by wall potentials of the forms   
$U_{\rm L}(x^{\rm H})=(x^{\rm H}+5\ell)^{-4}$ and  
$U_{\rm R}(x^{\rm H})=(x^{\rm H}-5\ell)^{-4}$.

The time evolution of the $i$-th Brownian particle is described by the 
one-dimensional Langevin equation 
\begin{eqnarray}
\gamma \dot{x}_i = -\frac{\partial U(x_i)}{\partial x_i} &+& f - 
\frac{\partial U_{\rm int}(x_i-x^{\rm H})}{\partial x_i}+\xi_i(t), \nonumber \\
\bra \xi_i(t)\xi_j(t')\ket &=& 2\gamma T\delta (t-t')\delta_{i,j}, 
\label{lan}  
\end{eqnarray}
and that of the Hamiltonian particle is described by 
\begin{eqnarray} 
m\dot{v}^{\rm H} &=&-\sum_{i=1}^{N}\frac{\partial U_{\rm int}(x_i-x^{\rm H})}
{\partial x^{\rm H}} - 
\frac{\partial U_{\rm L}(x^{\rm H})}{\partial x^{\rm H}}
- \frac{\partial U_{\rm R}(x^{\rm H})}{\partial x^{\rm H}}, \nonumber \\
\dot{x}^{\rm H}&=& v^{\rm H}.
\label{hami}
\end{eqnarray}
In  our numerical simulation, the velocity Verlet method was adopted to 
integrate the equation of motion (\ref{hami}) with a time step 
$\Delta t=5\times 10^{-5}$, and we used the parameter values $T=1$, 　
$\gamma=1$, $\ell=1$, $U_0=3$, $\varepsilon=1$, $m=1$, $r_{\rm c}=4$　and 
$0\le f \le 25$.  As to an initial condition,  $x^{\rm H}(0)=0$, 
$x_i(0)=i/2-10$ ($i=1,2,\cdots,20$)  and $v^{\rm H}(0)$ was chosen 
randomly according to a Gaussian distribution.

\begin{figure}
\begin{center}
\includegraphics[width=7cm]{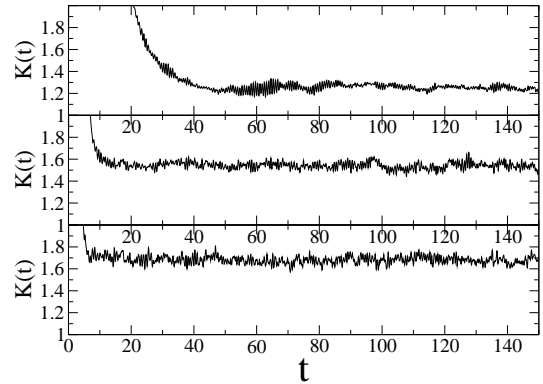}
\caption{$K(t)$  obtained using a moving thermostat as a function of 
time in the cases  $f=5$ (upper), $f=10$ (middle) and $f=15$ (lower). 　
These results were obtained from 5000 samples.}
\label{fig:ondo3}
\end{center}
\end{figure}

We define the velocity fluctuation of the Hamiltonian particle as 
\begin{equation}
K(t)\equiv m\left\{ \bra v^{\rm H}(t)^2\ket-\bra v^{\rm H}(t)\ket^2\right\}. 
\end{equation}
In Fig. \ref{fig:ondo2}, we plot $K(t)$ as  a function of time for $f=0$ 
and $f=10$ in the case of a stationary thermostat.  Next, we define the 
kinetic temperature as 
\begin{equation}
\bar{K} \equiv \lim_{t\to\infty} K(t). 
\end{equation}
In the case $f=0$, 
we find that $\bar{K}=1.00\pm 0.016$, which is equal to the temperature 
of the environment ($T=1$). In the case $f=10$, because the Brownian 
particles exhibit a non-zero average velocity maintained by $f$, 
significantly more heat flows into the Hamiltonian system than in the 
case $f=0$.  But in this case, as in the $f=0$ case, as $t$ increases, 
$K(t)$ approaches a constant value, with a kind of stationary behavior  
being established between the Hamiltonian system and the thermostat. In 
this case, we find $\bar{K}=6.22\pm 0.22$.  (We obtain $\bar{K}$ by 
averaging $K(t)$ over the intervals $t \in [60, 150]$ and $t\in [200, 300]$ 
for $f=10$ and $f= 0$, respectively.)

\begin{figure}
\begin{center}
\includegraphics[width=7cm]{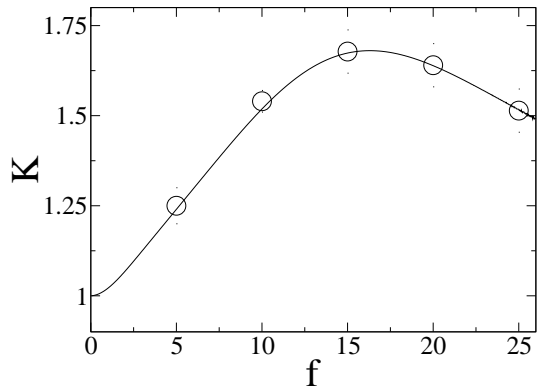}
\caption{The kinetic temperature $\bar{K}$ obtained using a moving 
thermostat as a function of $f$ (circles). The solid curve denotes the 
analytical solution of $\Theta(f)$ ($\equiv D(f)/\mu_{\rm d}(f)$) for 
the model (\ref{model0}) \cite{hs3a}. 
}
\label{fig:ondo4}
\end{center}
\end{figure}
%


{\it Moving thermostat}.   
Next, we consider the case of a moving thermostat. Specifically, we 
study the situation in which the thermostat moves at a constant 
speed of $v=-v_{\rm s}(f)$ relative to the Hamiltonian system,  where 
$v_{\rm s}(f)$ is the steady state velocity of the Brownian particles 
\cite{hs3a}.  With such a moving thermostat, the average velocity of 
each Brownian particle measured with respect to the spatial coordinate 
of the Hamiltonian system vanishes. Defining $y_i\equiv x_i-v_{\rm s}(f)t$, 
we can realize such a system by simply replacing $U_{\rm int}(x_i-x^{\rm H})$ 
with $U_{\rm int}(y_i-x^{\rm H})$ in Eqs. (\ref{lan}) and (\ref{hami}).

In Fig. \ref{fig:ondo3}, we plot $K(t)$ as a function of time in the 
cases $f=5$, $10$ and $15$. Note that in order to obtain these results 
for $K(t)$, we use the analytical solutions of $v_{\rm s}(f)$ for the model 
(\ref{model0}) \cite{value}.  From the data plotted in the graphs of 
Fig. \ref{fig:ondo3}, we find $\bar{K}=1.25\pm 0.05$ in the case $f=5$, 
$\bar{K}=1.54\pm 0.03$ in the case $f=10$, and $\bar{K}=1.68\pm 0.06$ in 
the case $f=15$.  Comparing the middle graph of Fig. \ref{fig:ondo3} with 
the lower graph of Fig. \ref{fig:ondo2}, both corresponding to the case 
$f=10$,  we find that the value of $\bar{K}$ obtained when using the 
moving thermostat differs significantly from that obtained when using the 
stationary thermostat. 
 
Now, let us compare the above results for $\bar{K}$ with the values of 
$\Theta$ for the thermostat. In Ref. \cite{hs3a}, from calculations of 
$D(f)$ and $\mu_{\rm d}(f)$, it was found that $D/\mu_{\rm d}=1.24$ in the 
case $f=5$, $D/\mu_{\rm d}=1.52$ in the case $f=10$, and $D/\mu_{\rm d}=
1.67$ in the case $f=15$ for the model (\ref{model0}). Because $T=1$,  
these values of $D/\mu_{\rm d}$ indicate  that the Einstein relation 
(\ref{ein}) does not hold for $f\ge 5$. Comparing these values with the 
values of $\bar{K}$ computed presently,  we find that the relation 
$\bar{K}=D/\mu_{\rm d}$ holds when $\varepsilon$, which represents the 
strength of the interaction between the thermostat and the Hamiltonian 
system, is sufficiently small \cite{eps}. This implies that the kinetic 
temperature 
of the Hamiltonian system is equal to the effective temperature $\Theta$ 
given by (\ref{the}) in the case that the Hamiltonian system is in contact 
with the moving thermostat.  In Fig. \ref{fig:ondo4},  we compare 
$\bar{K}$ with $\Theta$ ($\equiv D/\mu_{\rm d}$) for various values of 
$f$.  It is seen that the relation 
\begin{equation}
\bar{K}=\Theta
\label{kekka}
\end{equation}
holds, to the precision of the numerical computations.

\begin{figure}
\begin{center}
\includegraphics[width=7cm]{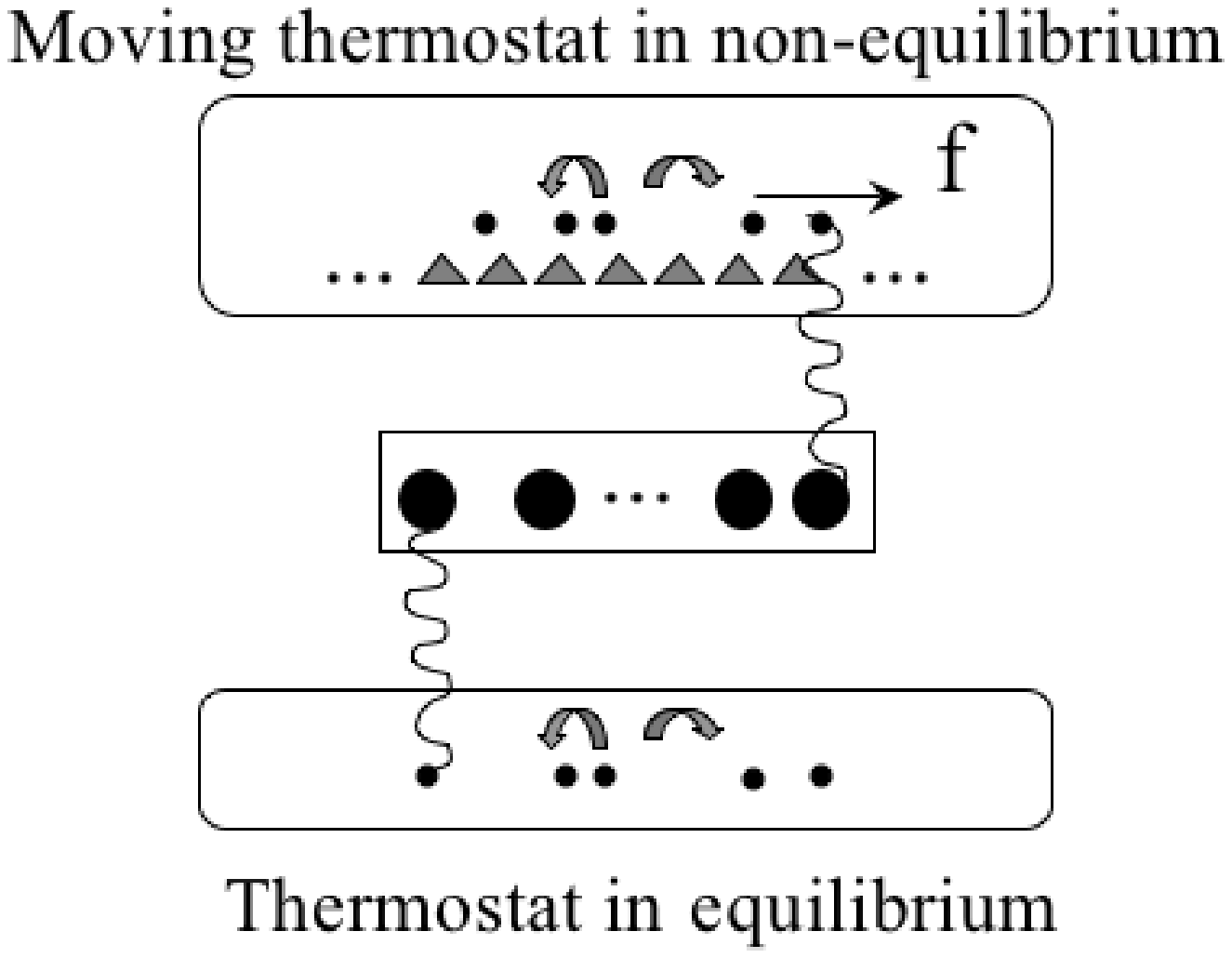}
\caption{Schematic depiction of the heat conduction system using 
the effective temperature.  
}
\label{fig:ondo5}
\end{center}
\end{figure}
%


{\it Interpretation of Eq. (\ref{kekka}).}    
In Ref. \cite{hs3a},   a large-scale description of the probability 
density for the model (\ref{model0}) was derived using a perturbation 
method, and it was found that $\Theta$ appears as a temperature in a  
Fokker-Planck equation of the coarse-grained  probability density. 
Then, in order to further investigate the physical properties of 
$\Theta$,  in Ref. \cite{hs5}, a coarse-grained description of the motion 
of a Brownian particle was derived by computing a finite time average of 
the Langevin equation, rather than analysing the probability density. 
This coarse-grained description is given by the equations 
$\Gamma (X_{n+1}-X_n)/\delta t  = F + \Xi_n$ and $\bra \Xi_n\Xi_m \ket 
\delta t=2\Gamma\Theta \delta_{m,n}$, where we have $F\equiv \Gamma 
v_{\rm s}$, $X_n\equiv x(t_n)$ and  $t_n\equiv n\delta t$ ($n=0,1,2,
\cdots$),  and the time interval $\delta t$ is chosen to be sufficiently 
longer than the characteristic time of the system. Here, $\Gamma$ and 
$F$ are uniquely determined as functions of the parameters that appear 
in the model (\ref{model0})  \cite{hs5}. Then, using the moving 
coordinates $Y_n\equiv X_n-v_{\rm s}t_n$, we can describe the large-scale 
motion of a Brownian particle by the equilibrium-form Langevin equation  
\begin{equation}
\Gamma  \frac{Y_{n+1}-Y_n}{\delta t}= \Xi_n. 
\label{sosika}
\end{equation}

In the present investigation, choosing the cut-off length of the 
interaction between the Brownian particles and the Hamiltonian particle, 
$r_{\rm c}$, to be sufficiently large, we considered the change in behavior 
of the system as we increase the number of the Brownian particles that 
interact with the Hamiltonian particle. In the case that there are many 
Brownian particles,  the Hamiltonian particle  moves slowly enough that 
its characteristic time is larger than $\delta t$. Because in this case, 
when we use the moving thermostat,  the motion of each Brownian particle 
is described by (\ref{sosika}), we obtain the result (\ref{kekka}). 
 
{\it Heat conduction.}  
As an application of the moving Langevin thermostat, we study the heat 
conduction system described below (see the schematic depiction in Fig. 
\ref{fig:ondo5}).  Here, a  one-dimensional Hamiltonian system consisting 
of $10$ particles is in contact with two thermostats, a moving, 
non-equilibrium thermostat of the type described above and an equilibrium 
thermostat.  Although the temperatures of the environments of both 
thermostats are set to  $T=1$, it is expected that a non-zero heat flux 
will be observed in the Hamiltonian system  because $\Theta$ (which 
differs from $T$) plays the role of the temperature in the moving thermostat.

\begin{figure}
\begin{center}
\includegraphics[width=7cm]{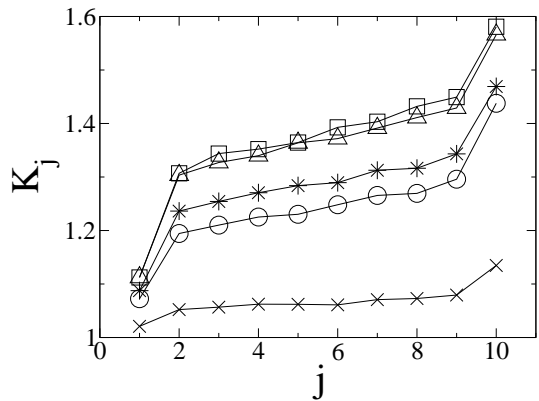}
\caption{$\bar{K}_j$ as a function of $j$ in the cases $f=5$ (pluses), 
$f=10$ (circles), $f=15$ (squares),  $f=20$ (triangles) and $f=25$ 
(asterisks). 
}
\label{fig:ondo6}
\end{center}
\end{figure}

Let  $x^{\rm H}_j$ be the position of the $j$-th Hamiltonian particle 
($j=1,\cdots 10$).  In our model, the $j$-th particle interacts only with 
its neighbors  (the $j\pm 1$-th particles) through the potential 
$U_{\rm int}^{\rm H}(x^{\rm H}_j-x^{\rm H}_{j\pm 1})=(1/2)(x^{\rm H}_j-x^{\rm H}
_{j\pm 1})^{2}+(10/4)(x^{\rm H}_j-x^{\rm H}_{j\pm 1})^{4}$.  Then, only the 
$1$-st Hamiltonian particle is in contact with the equilibrium thermostat, 
and only the $10$-th Hamiltonian particle is in contact with the moving 
thermostat.  

Defining $\bar{K}_j$ as the kinetic temperature of the $j$-th Hamiltonian 
particle, in Fig. \ref{fig:ondo6}, we plot $\bar{K}_j$.  It is seen that 
$K_j < K_{j+1}$. This is due to the relation 
$\Theta > T$.  Although $\Theta > T$ in our model, it has been reported 
that the case $\Theta < T$ can also be realized with an appropriate 
choice of the periodic potential $U(x_i)$ \cite{sasaki}.  This implies 
that we could control the direction of the heat flux by altering the  
functional form of $U(x_i)$.  

{\it Conclusion.}  
In this paper, we have investigated the use of a Langevin system in a 
NESS as a thermostat to establish the kinetic temperature of a Hamiltonian 
system.  Our main results consist of the relation (\ref{kekka}) and the 
data plotted in Fig. \ref{fig:ondo6},  both obtained with the use of the 
moving Langevin thermostat.  Because the physical relevance of effective 
temperatures  in NESS  \cite{hs3a,hs5}, glassy systems 
\cite{ckp,crisan,barrat}  and  biomolecules \cite{takano}  is not yet 
fully clarified,  we hope that our study sheds more light on it. 

\begin{acknowledgments}
We acknowledge S. Sasa and  H. Watanabe for discussions of this 
work and M. Otsuki for help with technical details.  We also thank 
G. Paquette for a critical reading of this article. This work was 
supported by grants from JSPS Research Fellowships for Young Scientists  
and the Ministry of Education, Science, Support and Culture of Japan.  
\end{acknowledgments}


\begin{thebibliography}{20}
\bibitem{value}
For the model (\ref{model0}), the values of $v_{\rm s}(f)$ are calculated 
to be $v_{\rm s}(5)=0.3840675$,  $v_{\rm s}(10)=2.022750$, $v_{\rm s}(15)=
6.025784$, $v_{\rm s}(20)=11.723569$ and 
$v_{\rm s}(25)=17.890192$, using Eq. (6) in Ref. \cite{hs3a}.
\bibitem{eps}
When $\varepsilon$ is large, the Langevin system is not the same 
as that studied in Ref. \cite{hs3a}. In this case, we can compute 
$D/\mu_{\rm d}$ numerically.
\end{thebibliography}
\end{document}